\documentclass[aps,prb,twocolumn,showpacs,amssymb]{revtex4}
\usepackage{bm}
\usepackage{graphicx}

\begin{document}

\title{$Sr_2RuO_4$: Broken Time-Reversal Symmetry in the Superconducting state}

\author{V.P. Mineev } 
\affiliation{Commissariat \`a l'Energie Atomique,
DSM/DRFMC/SPSMS 38054 Grenoble, France}

\date{\today}

\begin{abstract}
Using  a phenomenological two-fluid model we derive the Kerr rotation of the polarization direction of reflected light from the surface  of  a superconductor in a state breaking time-reversal symmetry. We argue that 
this effect  found recently in  the superconducting state of  $Sr_2RuO_4$ by Xia et al (Phys.Rev.Lett. {\bf 97}, 167002 (2006))  
originates from the spontaneous magnetization in this superconductor. The temperature and the frequency dependencies  of the effect  are established. It is shown that the effect is determined by one of two mechanisms depending on the frequency of light  that is larger or  smaller than the plasma frequency. The mechanism originating from the gradient of chemical potential 
created by the electric field of light penetrating into the skin layer and important at frequencies larger than the plasma frequency 
 was missed in the preliminary version of this paper (V.P.Mineev, cond-mat/0703624 V2).

\end{abstract}


\maketitle

\section{Introduction}

Xia et al \cite{Xia} have recently reported experimental evidence for the polar Kerr effect in the superconducting state of  $Sr_2RuO_4$ (rotation of polarization of reflected light). This observation made in the absence of external magnetic field demonstrates the time-reversal symmetry breaking 
in the superconducting state of this material.  The theoretical interpretation of the results has been given
first by the authors of  the original paper \cite{Xia} following a model for the Kerr effect in a metallic ferromagnet.
Subsequently a microscopic  approach to treat of this phenomenon in superconductors with spontaneous magnetization  \cite{Min} has been proposed by Yakovenko \cite{Yak}. Being  a fundamental method of checking of time reversal breaking, the polar Kerr effect in nonconventional superconductors  deserves a general phenomenological treatment  which is  developed  in the present article. 

Using the two-fluid approach generalized for a superconductor with time reversal breaking I derive
the Kerr angle rotation of the reflected light  polarization. 
In  correspondence with conditions of the experiment  \cite{Xia} the expression for the Kerr rotation is  found  at   high frequency  of light  when the latter is much larger than the quasiparticle scattering rate
$\omega\tau\gg1$ and exceeds the plasma frequency
$\omega>\omega_p$ in $Sr_2RuO_4$. 
The result consists of sum of two terms
\begin {equation}
\theta\cong-\frac{e^2 k_F}{\pi \hbar \omega}\frac{\Delta^2}{(\hbar\omega_p)^2}-\frac{n_n}{n_e\omega\tau}\frac{eH_s}{mc\omega}.
\label{e18x}
\end{equation}
Here $\Delta$ is the superconducting gap amplitude and $n_e$, $n_n=n_e-n_s$ are the total electron density and the density of the normal component correspondingly.
The first contribution proportional to square of  $\Delta$ originates from the gradient of chemical potential 
created by the electric field of light penetrating into the skin layer near  the superconductor 
surface. This term corresponds  to
the mechanism of Kerr rotation pointed out by Yakovenko \cite{Yak} but, unlike to that paper, it is inversely
proportional to the light frequency and  not to the third power of it.

The second term is due to a stationary magnetic field $H_s$ produced either by the superconducting domains with spontaneous magnetization or by the vortices trapped in the bulk of the superconductor  revealed
in $Sr_2RuO_4$ by scanning SQUID and Hall probe microscopy \cite{Dol, Bjorn, Kirtley}. The measurements in \cite{Xia}
has been performed in the zero field cooling and in the field cooling regimes. The superconducting signal is the difference between the measured Kerr angles at $T<T_c$ and  just above $T_c$ at the same field. 
So, in the field cooling regime $H_s$ has the meaning of difference between the actual fields near the surface of specimen at $T<T_c$ and at $T>T_c$.

 The Kerr effect described by the second term was considered in the preliminary version of this paper \cite{Min07} where the existence of the first term was 
missed.
The comparison of two terms demonstrates that the first contribution is always much larger than the second one. This fact is in accordance with an important observation done by Xia and co-authors   \cite{Xia} that the measured Kerr angle is independent of applied magnetic field. 

 The situation is changed in the frequency region $\omega<\omega_p$. Here the main contribution to the Kerr rotation originates of the stationary magnetic field $H_s$ but not from the electric field of light.
The measurements in this region can be used for the field $H_s$ determination.

\section{Kerr rotation in a superconducting state with spontaneous magnetization}

 We consider  linearly polarized light normally  (along the $z$-direction) incident from vacuum to the boundary of a medium with complex index of refraction
 \begin {equation}
 N=n+i\kappa
 \end{equation}
expressed through the diagonal part of complex conductivity $\sigma_{xx}=\sigma'_{xx}+i\sigma''_{xx}$
by means of the usual relations 
  \begin {equation}
n^2-\kappa^2=1-\frac{4\pi\sigma''_{xx}}{\omega},~~~~2n\kappa=\frac{4\pi\sigma'_{xx}}{\omega}.
\label{e2}
 \end{equation}
The light is reflected as elliptically polarized with the major axis rotated relative to the incident polarization by an amount \cite{Ben}
\begin {equation}
\theta=\frac{(1-n^2+\kappa^2)\Delta\kappa+2n\kappa\Delta n}{(1-n^2+\kappa^2)^2+(2n\kappa)^2},
\label{e3}
\end{equation} 
where 
\begin {equation}
\Delta n=n_+-n_-=-\frac{4\pi}{\omega}\frac{n\sigma'_{xy}+\kappa\sigma''_{xy}}{n^2+\kappa^2},
\label{e4}
\end{equation} 
\begin {equation}
\Delta\kappa=\kappa_+-\kappa_-=-\frac{4\pi}{\omega}\frac{n\sigma''_{xy}-\kappa\sigma'_{xy}}{n^2+\kappa^2}
\label{e5}
\end{equation} 
are the difference in the real (imaginary) parts of circularly polarized lights  with the opposite polarization and $\sigma_{xy}=\sigma'_{xy}+i\sigma''_{xy}$ is the complex off diagonal component of the conductivity tensor.

The diagonal and the off diagonal component of conductivity can be found in the frame of two fluid phenomenology
where the current density consists of the sum of densities of normal and superfluid current 
\begin {equation}
{\bf j}={\bf j}_n+{\bf j}_s.
\end{equation}
The normal current density is determined by the standard equation of motion
\begin {equation}
\left (-i\omega+\frac{1}{\tau}\right ){\bf j}_n=\frac{e^2n_n}{m}\left({\bf E}+
 \frac{{\bf j}_n\times{\bf H}_s}{en_nc}\right ).
 \label{e5a}
\end{equation}
The superfluid current density in a superconductor with spontaneous magnetic moment has the form \cite{Muz,Bal}
\begin {equation}
{\bf j}_s=en_s({\bf v}_s-{\bf v}_n)+ \frac{e\hbar A}{4m}\nabla n_e\times\hat{\bf m}.
\end{equation}
The unit vector $\hat{\bf m}$ is the direction of spontaneous magnetization. It has  the space variations near the superconducting domain walls  and the specimen surface \cite{Vol} then the current expression has a more general form.  Here we  assume that $\hat{\bf m}$ is pinned to a crystallographic direction (in the case of $Sr_2RuO_4$ to the tetragonal axis). The gradient of electron density arises near the specimen boundaries and also due to an electric field ${\bf E}$ of light penetrating in the skin layer near the  superconductor  surface.  Only the latter is  important for the complex conductivity 
determination. So, one can rewrite the expression for the current as
\begin {equation}
{\bf j}_s=en_s({\bf v}_s-{\bf v}_n)+\frac{e^2\hbar AN_0}{4m}{\bf E}\times\hat{\bf m},
\label{e55}
\end{equation}
here $N_0$ is the electron density of states. In the static case  the coefficient A is found \cite{Bal} equal to $n_s/n_e$. So, $A=1$
 at $T=0$ and in 2D case when $N_0=m/\pi\hbar^2$ we have for the Hall conductivity 
 $\sigma_{xy}=e^2/2h$, as it was pointed out in
\cite{Vol88}. In the general case of the finite frequency one can calculate the coefficient $A$ as the 
transversal current response to an alternating field. Here we shall use the value 
$A\propto i\Delta^2/(\hbar\omega)^2$ found in the high frequency  limit $\hbar\omega\gg\Delta$ in the paper \cite{Yak}. This expression is valid up to some numerical coefficient of the order of unity depending on particular form of the order parameter.

The part of superfluid current  (\ref{e55}) $\tilde{\bf j}_s={\bf j}_s-e^2\hbar AN_0({\bf E}\times\hat{\bf m})/4m$ proportional to the velocities difference
should obey the London equation of motion
\begin {equation}
-i\omega~\tilde{\bf j}_s=\frac{e^2n_s}{m}\left({\bf E}+
 \frac{\tilde{\bf j}_s\times{\bf H}_s}{en_sc}\right ).
 \label{e5b}
\end{equation}

The strong inequality $\omega_s=eH_s/mc=1.76\times 10^7 H_s(Oe)\ll\omega$ is obviously fulfilled for an infrared region of frequency at any reasonable value of magnetic field. 
Hence, the solutions of the equations (\ref{e5a}) and (\ref{e5b}) are 
\begin {equation}
{\bf j}_n=\frac{e^2n_n\tau}{m(1-i\omega\tau)}\left({\bf E}+
\frac{e\tau({\bf E}\times{\bf H}_s)}{mc(1-i\omega\tau)} 
 \right ),
 \label{e11}
\end{equation}
\begin {equation}
{\bf j}_s=\frac{ie^2k_F}{4\pi^2\hbar}\frac{ \Delta^2} {(\hbar\omega)^2}({\bf E}\times\hat{\bf m})-\frac{e^2n_s}{m i\omega}\left({\bf E}-
\frac{e({\bf E}\times{\bf H}_s)}{mc~ i\omega} 
 \right ).
 \label{e12}
\end{equation}
Here the value of  3D density of states $N_0=mk_F/\pi^2\hbar^2$  was substituted.

The light frequency used in the experiment \cite{Xia} is $\omega\sim 10^{15} rad/sec$
and scattering time  $\tau\sim 10^{-11} sec $ (for a review of experimental data see \cite{Mack}).  Hence in what follows we shall discuss the frequency region $\omega \tau\gg1$. Then  from  
(\ref{e11})-(\ref{e12}) choosing ${\bf E}\parallel\hat y$, ${\bf H}_s\parallel\hat{\bf m}\parallel\hat z$ we obtain

\begin {equation}
\sigma'_{xx}
\cong\frac{\omega_p^2}{4\pi\omega}\frac{n_n}{n_e\omega\tau },
\end{equation}
\begin {equation}
\sigma''_{xx}
\cong\frac{\omega_p^2}{4\pi\omega},
\end{equation}
\begin {equation}
\sigma'_{xy}
\cong-\frac{eH_s}{4\pi mc}\frac{\omega_p^2}{\omega^2},
\label{e13a}
\end{equation}
\begin {equation}
\sigma''_{xy}\cong\frac{e^2k_F}{4\pi^2\hbar}\frac{ \Delta^2} {(\hbar\omega)^2}+
\frac{eH_s}{2\pi mc}\frac{\omega_p^2}{\omega^2}\frac{n_n}{n_e\omega\tau}.
\label{e14a}
\end{equation}
We are working in frame of one band approach and use the notation $\omega_p=\sqrt{4\pi n_e e^2/m}$ for the plasma frequency. Actually $Sr_2RuO_4$ is three-band metal and each of three bands has its own plasma frequency. 
So, in what follows,  the inequality $\omega>\omega_p$
means that the frequency exceeds the largest plasma frequency of the metal. The opposite inequality 
$\omega<\omega_p$ means that we consider the frequencies which are smaller than the smallest plasma frequency of the metal. If the actual frequency is in an interval between the different band plasma frequencies then the Kerr rotation has some value,  intermediate between its values
at $\omega<\omega_p$ and at $\omega>\omega_p$ written below.
The average plasma frequency 
 in the $ab$ plane of $Sr_2RuO_4$ is somewhat smaller than $10^{15} rad/sec $ \cite{Kats}.

The experiment \cite{Xia} has been performed under conditions of  good reflectivity $R\approx 0.6$. The latter at $\omega\tau\gg1$ takes place  \cite{Ziman} in the frequency region 
$\omega>\omega_p$ where the solution of eqns (\ref{e2}) has the form
\begin {equation}
n\cong\frac{\sqrt{\omega^2-\omega_p^2}}{\omega},~~~~~
\kappa\cong\frac{n_n}{2n_e\omega^2\tau}\frac{\omega_p^2}{\sqrt{\omega^2-\omega_p^2}}.
\label{e17}
\end{equation}
So, taking into account that $n<1$ and $\kappa\ll 1$ we obtain for the angle of rotation of the polarization
\begin {equation}
\theta\cong\frac{4\pi}{\omega}\left (\frac{\sigma''_{xy}}{n(n^2-1)}+
\frac{\kappa(1-3n^2)\sigma'_{xy}}{n^2(n^2-1)^2}\right ).
\label{e89}
\end{equation}
The first term here corresponds to the initial formula for $\theta$ used  by Yakovenko \cite{Yak}.
He has substituted $n(n^2-1)=3$, that is $n\approx 1.7$. This magnitude  of $n$ is  in principle suitable
at light frequencies $\omega<\omega_p$ for not so large values of the product $\omega\tau$.
However,  in this frequency region, corresponding to almost ideal reflectivity, the Eqn. (\ref{e89}) does not work.
Thus, at 
$\omega>\omega_p$, using Eqns. (\ref{e17}), we come to
\begin {equation}
\theta\cong-\frac{e^2k_F}{\pi\hbar\omega}\frac{ \Delta^2} {(\hbar\omega_p)^2}\frac{1}{\left(1-\frac{\omega_p^2}{\omega^2}\right)^{1/2}}-\frac{n_n}{n_e\omega\tau}\frac{eH_s}{2mc\omega}
\frac{2-\frac{\omega_p^2}{\omega^2}}
{\left(1-\frac{\omega_p^2}{\omega^2}\right)^{3/2}}.
\label{e18}
\end{equation}
In negligence of the ratio $\omega_p^2/\omega^2$ we come to the simplified formula (\ref{e18x})
written in the Introducion.
The substitution of numerical values $\omega\approx\omega_p\approx 10^{15}rad/sec$ and $k_F\approx 10^7 cm^{-1}$ gives for the first term 
\begin {equation}
|\theta|\approx 10^{-8}\frac{\Delta^2}{T_c^2}~rad,
\label{e20}
\end{equation}
that appears to be a reasonable estimation of the Kerr angles observed in the paper \cite{Xia}.
The numerical value of 
 the second term contribution has the magnitude $10^{-12} H_s(Oe)rad$.
 So, this term is negligibly small in comparison with the first one at any reasonable values of $H_s$. 
 Hence, the superconducting  signal should be roughly the same  in the zero field cooling and the field cooling measurements.
 
 The situation is drastically changed in the  region of perfect reflectivity $R\cong 1$  at $\omega<\omega_p $.
 Here the solution of equations (\ref{e2}) is 
\begin {equation}
n\cong\frac{n_n}{2n_e\omega^2\tau}\frac{\omega_p^2}{\sqrt{\omega_p^2-\omega^2}},~~~~~
\kappa\cong\frac{\sqrt{\omega_p^2-\omega^2}}{\omega}.
\label{e15}
\end{equation}
and for the angle of rotation of the polarization we obtain
\begin {equation}
\theta\cong
-\frac{n_n}{n_e\omega\tau}\frac{e^2k_F}{\pi\hbar\omega}\frac{\Delta^2}{(\hbar\omega_p)^2}\frac{\frac{3\omega_p^2}{2\omega^2}-1}{\left(\frac{\omega_p^2}{\omega^2}-1\right)^{3/2}}-\frac{eH_s}{mc\omega}\frac{1}{\sqrt{\frac{\omega_p^2}{\omega^2}-1}}.
\label{e16}
\end{equation}
The comparison of two terms shows that in this frequency region the second term proportional to $H_s$
is much more important. The measurements of Kerr angle at $\omega<\omega_p$ can serve for  the field $H_s$ determination. The problem, however, is  complicated because, as it is in the usual Hall effect, the electron and the hole bands give  the different sign contributions to this term. 

The field $H_s$ originating of spontaneous magnetization is of the order of the lower critical field \cite{Min,Vol}.  It will,  in fact,  be strongly suppressed due to cancellation of the signal from the  domains  with opposite directions of magnetization and due to the effect of magnetic field shielding at distances larger than the London penetration depth from the domain boundaries.
Hence, the determination of spontaneous magnetization by the Kerr angle measurements is quite problematic. On the other hand, the Kerr angle measurements in the field cooling regime at $\omega<\omega_p$ probably can reveal the signal field dependence arising due to the difference in the actual fields near the specimen surface in the superconducting $(T<T_c)$ and and in the normal $(T>T_c)$ states caused by the field repulsion weakened by the penetration of vortices.

\section{Conclusion}

The proposed theoretical treatment of the Kerr rotation is definitely in favor of an interpretation of the experimental results \cite{Xia} in terms of spontaneous time reversal breaking in the superconducting $Sr_2RuO_4$. On the other hand,  a superconducting state possessing spontaneous magnetization is inevitably described by multicomponent order parameter \cite{Min}. 
In a tetragonal  crystal the  two-component superconducting states 
 either with singlet or with triplet pairing are admissible \cite{Min}. 
The specific properties for these states are: (i) the basal plane anisotropy of the upper critical field \cite{Gor} and (ii) the  splitting of the phase transition to superconducting state
under magnetic field oriented in the  basal plane \cite{Min} . 
Due to the symmetry reason the properties take place also in multiband superconductors.
In application to $Sr_2RuO_4$ (although in a single band model) these properties  have been thoroughly theoretically investigated  by Kaur, Agterberg and Kusunose \cite{Kaur}. The unavoidable contradiction has been demonstrated: one particular choice of  two-component order parameter is appropriate for the elimination of the basal plane upper critical field anisotropy but at the same time a considerable phase transition splitting occurs. 
Vice versa, another particular choice of the order parameter almost eliminates the phase transition splitting for the  one particular field direction but keeps the basal plane upper critical field anisotropy. 
Both of these properties should manifest themselves below the critical temperature \cite{Min} but untill now there is no experimental evidence for this. The in-plane  anisotropy of the upper critical field has been observed only at low temperatures \cite{Mao} where it  is quite well known phenomenon for any type of superconductivity
originating from the Fermi surface anisotropy.  Recent detailed magnetization measurements 
\cite{Tenya}  have shown
an unusual magnetic response  looking  not convincing enough to be the  trace of  an additional phase transition. Peculiar features are revealed in low temperature behavior of susceptibility and specific heat under in-plane magnetic field \cite{Yaguchi,Deguchi}. The theoretical explanation of them has been recently proposed in paper \cite{Machida}.

Thus, several experimental observations are incompatible with multicomponent order parameter structure dictated by Kerr rotation  \cite{Xia}, muon spin rotation \cite{Luke} and phase sensitive Josephson  \cite{Liu,Kid} measurements manifesting the spontaneous time-reversal breaking.
So, the problems relating to the superconducting $Sr_2RuO_4$ still exist.

\section*{ACKNOWLEDGMENTS}

The author is indebted to V.Egorov and B.Spivak for valuable discussions and to A. Lebed', K.Hasselbach, 
K. Machida, G. Luke and A. Kapitulnik for the interest to this work. I would also like to thank  D.Maude for the help in the manuscript preparation.

\end{document}